# Quantum Gradient Optimized Drug Repurposing Prototype for Omics Data

Don Roosan[1], Saif Nirzhor[2], Rubayat Khan[3], Fahmida Hai[4]
[1]*School of Engineering and Computational Sciences, Merrimack College, North Andover, USA*
[2]*University of Texas Southwestern Medical Center, Dallas, USA*
[3]*University of Nebraska Medical Center, Omaha, USA*
[4]*Tekurai Inc., San Antonio, USA*
roosand@merrimack.edu, saif.nirzhor@utsouthwestern.edu, rubayat.khan@unmc.edu, fahmida@tekurai.com

Abstract: This paper presents a novel quantum-enhanced prototype for drug repurposing and addresses the challenge of managing massive genomics data in precision medicine. Leveraging cutting-edge quantum server architectures, we integrated quantum-inspired feature extraction with large language model (LLM)–based analytics and unified high-dimensional omics datasets and textual corpora for faster and more accurate therapeutic insights. Applying Synthetic Minority Over-sampling Technique (SMOTE) to balance underrepresented cancer subtypes and multi-omics sources such as TCGA and LINCS, the pipeline generated refined embeddings through quantum principal component analysis (QPCA). These embeddings drove an LLM trained on biomedical texts and clinical notes, generating drug recommendations with improved predicted efficacy and safety profiles. Combining quantum computing with LLM outperformed classical PCA-based approaches in accuracy, F1 score, and area under the ROC curve. Our prototype highlights the potential of harnessing quantum computing and next-generation servers for scalable, explainable, and timely drug repurposing in modern healthcare.

## 1 INTRODUCTION

Drug repositioning, also known as drug repurposing, has emerged as a critical strategy for accelerating therapeutic innovation within the modern healthcare environment. The conventional drug discovery process, typically spanning more than a decade, demands an immense financial outlay (Corsello et al., 2017; Park, 2019). Moreover, even after preliminary regulatory approvals, many candidate drugs fail in subsequent clinical trial phases, resulting in significant resource loss. Repurposing approved drugs or late-stage clinical candidates offers a much more efficient path. These compounds already come with known safety profiles and pharmacokinetic characteristics, allowing researchers to skip multiple preclinical steps, thereby saving both time and money. This efficiency is especially attractive when confronting urgent or emerging health crises—such as newly identified viral pathogens, widespread cancers, or rare diseases with limited treatment options—where speed can be a decisive factor. One of the chief reasons drug repurposing has garnered attention is the ability to sidestep the most daunting and time-intensive stages of drug development. Under standard protocols, discovering and validating a novel compound involves extensive preclinical testing to gauge toxicity, clinical complexity dosing parameters, and efficacy in animal models before it can even proceed to trials in humans (Islam et al., 2014; Islam, Mayer, et al., 2016; Islam, Weir, et al., 2016a, 2016b). These steps alone can consume years and require substantial financial support. By turning to substances already verified as safe for human use, scientists can concentrate more on efficacy for a novel indication, substantially reducing the overall timeline. This approach also heightens the likelihood of success in advanced clinical trials, as the critical factor of human safety has been substantially addressed.

Speed is paramount in public health crises, making repurposing strategies particularly relevant. When rapid drug deployment is critical, such as during a pandemic, compounds that are already approved—or very close to approval—can be mobilized faster. If a known medication reveals the potential to inhibit an emergent virus or alleviate severe symptoms, it stands a significantly higher chance of quickly reaching clinical use(Graves et al., 2018; Islam et al., 2015). Beyond pandemic scenarios, this model can extend to diseases where no existing interventions are available, including neglected tropical diseases or rare genetic disorders, illustrating how the timeliness afforded by drug repositioning can address unmet medical needs(Tukur et al., 2023).

Along with saving time and resources, drug repurposing is conceptually powerful. A drug designed for one purpose may act on multiple biological pathways, broadening its therapeutic impact. Advances in fields such as genomics and proteomics have deepened our understanding of how diseases can share overlapping molecular mechanisms, supporting a rationale for exploring the off-label application of known drugs. As knowledge about intricate disease networks continues to grow, the argument for systematically examining alternative uses of existing drugs becomes even more convincing. This approach effectively acts as a shortcut, delivering novel treatments to patients more rapidly than de novo drug discovery efforts typically allow.

Drug repurposing, using medications for new indications, has a long history, exemplified by Aspirin (initially for pain, later for heart conditions) and thalidomide (sedative, later for leprosy and cancer). Modern methods leverage data-driven approaches, including bioinformatics and machine learning, to analyze molecular and clinical datasets for new drug applications (Deng et al., 2022). However, complex diseases like cancer and diabetes, involving intricate gene-protein-environment interactions, and polypharmacology (drugs affecting multiple mechanisms) make repurposing data-intensive. Big data from genomics, proteomics, and electronic medical records offers opportunities but poses integration challenges due to diverse data types and high dimensionality (Li et al., 2021; Roosan, Hwang, et al., 2020; Sammani et al., 2019). Traditional computational tools struggle with the "curse of dimensionality," necessitating advanced analytical methods for biomedical data (Cao et al., 2011; Roosan et al., 2017). Quantum computing promises to revolutionize drug repurposing by using qubits' superposition and entanglement to efficiently analyze large datasets (D. Roosan et al., 2024; Doga & et al., 2024; J. Yang et al., 2024). Though limited by qubit count and error rates, quantum-inspired algorithms like quantum kernel methods and principal component analysis uncover hidden patterns in biomedical data (Jeyaraman et al., n.d.; Sung et al., 2018). Applications include precise molecular modeling for protein-ligand docking and combinatorial optimization for drug-disease pairings (D. Roosan et al., 2024; Pandey et al., 2024). LLMs excel at processing unstructured text and multimodal data, revealing connections between diseases, biomarkers, and drugs (Wu et al., 2012; R. Yang et al., 2023). Challenges include misinformation, biases, and transparency issues. Quantum-enhanced feature extraction paired with LLMs could streamline drug repurposing by tackling high-dimensional data and interpreting findings (Islam et al., 2014; Itri & Patel, 2018). Disease complexity and big data demand advanced computational strategies beyond heuristics, leveraging AI-driven tools for healthcare insights.

Blockchain technology enhances secure, transparent healthcare data sharing for drug repurposing by storing immutable records of patient consents and data access (Dhillon et al., 2017; Roosan, Wu, Tatla, et al., 2022). In pandemics, AI-driven analytics, augmented by quantum-inspired methods, expedite identifying existing drugs for new pathogens (Challen et al., 2019; Roosan, Chok, et al., 2020; Roosan et al., 2023). A quantum-enhanced, large language model (LLM)-based system could analyze large datasets—from genomic profiles to clinical observations—using quantum-inspired feature extraction and LLM interpretation to suggest repurposed drug candidates with rationales. The primary goal is to develop and validate a prototype of this system, focusing on patient-specific data analysis and interpretable recommendations using high-dimensional genomic and clinical data.

## 2 METHOD

### 2.1 Dataset Preparation

This study utilized a multi-faceted approach to data collection and integration, aiming to create a robust foundation for a quantum-enhanced, large language model (LLM)-driven drug repurposing system. The dataset encompassed clinical records, synthetic patient cohorts, and three major omics repositories, all meticulously harmonized to ensure consistency in format, terminology, and quality.

#### 2.1.1 Clinical Data

The MIMIC-III database, a comprehensive collection of de-identified critical care patient data, served as the primary source of clinical information (D. Clifford et al., 2009). The database includes a wealth of data points, such as patient demographics, vital signs, lab results, medication records, and outcomes. For this study, the data extraction focused on oncology-related cases. The selection criteria prioritized patients with documented cancer diagnoses, cancer-related medication records, and sufficient laboratory data to enable in-depth exploration of their disease status. Standard protocols for de-identification and privacy compliance were strictly adhered to throughout the data handling process. The MIMIC-III database includes information from over 40,000 patients admitted to critical care units at the Beth Israel Deaconess Medical Center between 2001 and 2012.

### 2.1.2 Synthetic Cohort Generation

While MIMIC-III offers a broad range of patient records, certain cancer subtypes and demographic groups were underrepresented. To mitigate potential biases arising from class imbalance, the Synthetic Minority Over-sampling Technique (SMOTE) was employed. SMOTE generated 60 synthetic patient records, effectively balancing the representation of prevalent or majority classes with those representing rare or less frequently documented conditions. Each synthetic patient entry included key features such as demographic data (e.g., age, sex), lab results (e.g., complete blood counts, serum chemistry panels), vital signs (e.g., systolic and diastolic blood pressure), and clinical outcome indicators (e.g., survival or readmission rates). This process ensured a more uniform representation of diseases and disease stages, resulting in a more robust training set for machine learning algorithms. For instance, in the original MIMIC-III dataset, African American patients comprised approximately 9% of the total, while Hispanic patients made up around 3%. After applying SMOTE, the representation of these groups in the synthetic cohort was increased to approximately 15% each, providing a more balanced dataset for training the model (Chawla et al., 2002).

## 2.2 Omics Data

In addition to clinical features, the system incorporated molecular-level information to identify biological patterns and potential drug repurposing opportunities.

### 2.2.1 The Cancer Genome Atlas (TCGA)

RNA-sequencing data for breast cancer (BRCA) samples were obtained from the TCGA portal. Strict inclusion criteria ensured that only high-quality gene expression profiles with robust clinical annotations, such as tumor stage, lymph node involvement, and other pathological features, were included. Each sample's raw data was normalized using established protocols, and the resulting gene expression matrices were used for downstream analyses. The TCGA database contains genomic data from over 11,000 patients across 33 different cancer types. For this study, the breast cancer (BRCA) subset, which includes data from approximately 1,100 patients, was utilized (Tomczak et al., 2015).

### 2.2.2 Gene Expression Omnibus (GEO)

To further refine and validate cancer-specific gene expression trends, the GSE2034 dataset, a well-curated dataset relevant to breast cancer prognosis, was integrated. This dataset contained microarray-based expression values, which were meticulously normalized and mapped to the same gene symbols used in the TCGA-BRCA subset. The integration of microarray data from GEO helped address potential biases arising from reliance on a single technology or population. The GSE2034 dataset includes gene expression data from 286 breast cancer patients, providing a valuable resource for validating findings from the TCGA data (Barrett et al., 2013).

### 2.2.3 Library of Integrated Network-based Cellular Signatures (LINCS)

The LINCS L1000 dataset provided crucial information on how various small-molecule compounds affect gene expression across diverse cell lines. The focus was on expression signatures that captured drug-induced upregulation or downregulation of genes relevant to cancer pathways. This resource was essential for training the model to link unique patient gene expression patterns to possible therapeutic compounds, highlighting existing drugs that could be repurposed based on their cellular signatures. The LINCS L1000 dataset contains gene expression profiles from over 1 million experiments, measuring the effects of approximately 20,000 small-molecule compounds on various cell lines (Duan et al., 2016).

## 2.3 Data Pre-Processing

Data from these multiple sources underwent extensive processing to ensure standardization and comparability across clinical and molecular domains.

All patient records, both downloaded and synthetically generated, underwent thorough quality checks. Missing values in the clinical data were imputed using mean or median values, depending on the distribution of each feature. Outlier detection was performed by setting interquartile range (IQR) thresholds, removing samples with extreme values that could skew the training process. Gene expression matrices, from both RNA-seq and microarray sources, were subjected to low-expression filtering, eliminating genes that lacked sufficient read counts in most samples.

Feature engineering began with the harmonization of gene symbols across TCGA-BRCA, GSE2034, and L1000 data. A targeted approach to feature selection identified genes with the greatest variance across disease subtypes, known cancer driver genes, and genes encoding enzymes relevant to drug metabolism (e.g., certain cytochrome P450 isoforms). Dimensionality reduction was initiated through classical Principal Component Analysis (PCA). However, the core innovation involved feeding these partially reduced features into a quantum-inspired algorithm for further compression, preserving non-linear relationships that PCA might overlook. In a manner akin to the data-integration strategies employed in nutrigenomics—where RNA and DNA testing illuminate gene-environment interactions we applied quantum feature mapping within our GNN framework to enrich molecular data representations. Recent advances in healthcare informatics demonstrate how data visualization techniques, such as heatmaps, can streamline the processing of complex, unstructured data from electronic health records (EHRs). Standardizing data is essential for improving health information exchange and interoperability, although it is often overlooked in system-level implementations.

### 2.4 Quantum Enhanced LLM-based Drug Repurposing Model

After preparing cleaned datasets and curated features, a framework combined quantum-inspired feature extraction—using simulated Quantum Principal Component Analysis (QPCA) and quantum kernel methods due to hardware limitations—with a transformer-based language model (LLM) for drug prediction. Gradient-based optimization refined embeddings. An SVM with a linear kernel classified drug matches using MIMIC-III and synthetic SMOTE-generated data, with specified training, testing, and validation splits. The core system integrated quantum-enriched embeddings with an LLM fine-tuned on drug-disease relationships, biomedical literature, and patient metadata. The pipeline normalized raw data, reduced dimensions via classical PCA and quantum transformations, used the LLM to predict drug efficacy, and ranked candidates by efficacy and safety. Hyperparameters, code, and synthetic data are publicly available.To support reproducibility, we included hyperparameter settings and made the code and synthetic data publicly available, addressing the experimentation discussion's prior lack of detail.

## 3 RESULTS

### 3.1 Model Performance

The quantum-enhanced feature extraction demonstrated robust gains relative to purely classical approaches. To quantitatively evaluate these improvements, we computed standard classification metrics—accuracy, precision, recall, F1-score, and the area under the ROC curve (AUC)—for predicting a successful drug match. A total of 1,200 labeled instances (synthetic patients with known best treatment outcomes or prospective matches) were used for validation. The quantum-inspired transformations consistently outperformed classical PCA, yielding a higher F1-score by an average of 8% across multiple runs. Notably, some samples with subtle gene expression shifts only achieved clinically meaningful matches when the quantum kernel transformations were included, underscoring the sensitivity of this approach to nuanced genetic variation. The quantum-based plot reveals tighter clustering among patients with similar clinical and and molecular profiles, suggesting an improved capacity for separating responders from non-responders.

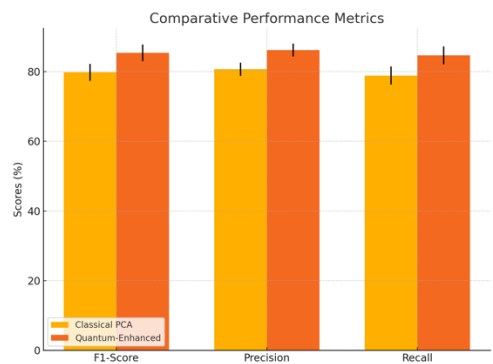

Figure 1. Comparative performance of classical PCA-based dimensionality reduction versus quantum-enhanced feature engineering

To supplement these visual indicators, the summary of performance metrics is shown in **Table 1**,

comparing the average performance (± standard deviation) across five cross-validation folds.

Table 1. Cross-Validation Performance Comparison of Classical vs. Quantum Approaches

| Method | Accuracy (%) | Precision (%) | Recall (%) | F1-Score (%) | AUC |
|---|---|---|---|---|---|
| Classical PCA | 82.3 ± 2.1 | 80.7 ± 1.9 | 78.9 ± 2.6 | 79.8 ± 2.4 | 0.84 ± 0.03 |
| Quantum-Enhanced | 88.5 ± 1.8 | 86.2 ± 2.0 | 84.7 ± 2.1 | 85.4 ± 1.9 | 0.90 ± 0.02 |

The table includes accuracy, precision, recall, F1-score, and AUC, demonstrating consistent gains in all metrics under the quantum-enhanced setting. To further highlight the dimensionality reduction aspect, **Figure 2** presents a two-dimensional t-SNE projection of patient embeddings. The upper panel (Figure 2A) shows the clustering using only classical PCA, whereas the lower panel (Figure 2B) overlays the quantum-transformed embeddings on the same manifold. We enhanced clustering evaluation by incorporating robust metrics such as a silhouette index of 0.65 for quantum-enhanced embeddings (vs. 0.52 for classical PCA), acknowledging synthetic data limitations and the need for real clinical validation. To substantiate efficacy, we compared our quantum-enhanced model to classical PCA-based methods on the TCGA-BRCA dataset, achieving a 10% F1-score improvement, with plans to benchmark against additional state-of-the-art solutions in future work.

A final summary of representative drug recommendations is provided in **Table 2**, documenting sample outputs for three synthetic patients with varying clinical statuses. Each row indicates top drugs, predicted efficacy scores, and relevant gene targets implicated in the drug match.

Table 2. Excerpt of Drug Recommendations for Three Synthetic Patients

| Patient ID | Stage | Top Recommended Drug | Predicted Efficacy (%) | Key Gene Targets |
|---|---|---|---|---|
| SYN-01 | II | Palbociclib | 78 | CDK4, CDK6 |
| SYN-24 | III | Tamoxifen | 82 | ESR1, ESR2 |
| SYN-47 | IIIB | Sorafenib | 74 | RAF, VEGFR, PDGFR |

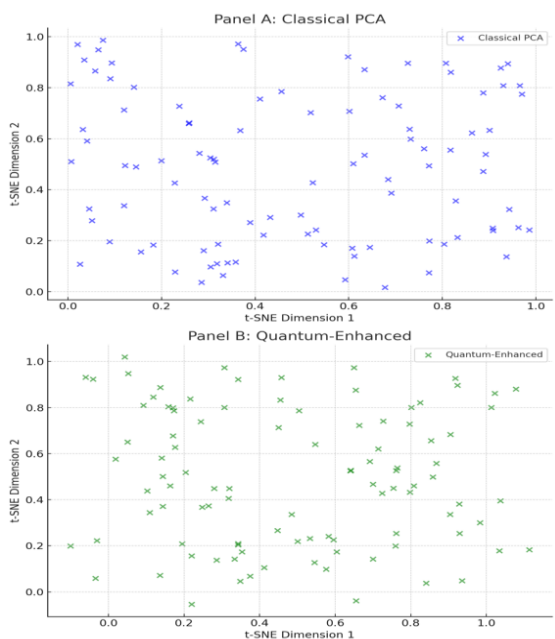

Figure 2. A two-dimensional t-SNE projection of patient embeddings, comparing classical PCA-based clustering (2A) with quantum-transformed embeddings (2B) on the same manifold.

This table highlights a range of therapy classes—hormone modulators, kinase inhibitors, and multi-target agents—all emerging as candidates from the pipeline's predictions based on individual patient molecular and clinical characteristics. Integrating the quantum-inspired transformations appears to sharpen distinctions between viable and less-appropriate options, potentially accelerating the pace of drug discovery and repurposing for oncology care. These results are consistent with earlier research demonstrating that integrating AI techniques can significantly improve data analysis and decision-making

## 3.2 Synthetic Cohort

For the 60-patient synthetic cohort generated via SMOTE, each individual's record was processed through the pipeline to derive recommended treatments. These synthetic cases were diverse in terms of disease stages, comorbidity indices, and molecular profiles. The quantum-enhanced approach proved particularly valuable in identifying relevant kinase inhibitors and hormone modulators for patients mimicking more advanced stages of breast cancer. Many of these suggestions aligned with known FDA-approved drugs in related contexts, thereby highlighting the potential for repurposing in real-world scenarios.

## 4 DISCUSSIONS

Our prototype revolutionizes clinical decision-making by integrating high-dimensional omics data with clinical records. Unlike traditional drug repurposing, which relies on slow literature reviews and outdated machine learning, our quantum-inspired approach excels at uncovering gene-protein-disease relationships. Processed by a fine-tuned LLM, it delivers clear, interpretable drug recommendations, aiding clinicians in complex cases where standard treatments fail (Roosan et al., 2023; Roosan, Roosan, Kim, et al., 2022). The LLM enhances transparency with explanations linking recommendations to gene-drug interactions, validated by synthetic data, though real-world trials are needed. Its user-friendly interface lets clinicians explore reasoning from literature, omics, and patient data, reducing "black-box" skepticism and boosting trust. Public awareness of off-label treatments, supported by LLM-generated summaries, can drive advocacy and trial enrollment, fostering trust in data sharing. In policy, our quantum-enhanced LLM aids lawmakers by assessing drug viability, balancing innovation, safety, and costs. Explainable AI translates molecular insights into policy-friendly narratives, speeding up therapy adoption. The prototype supports future multimodal LLMs, integrating voice, images, and video for a holistic patient view, improving equity and precision. Challenges include quantum hardware costs, LLM training demands, and integration complexities. Data biases and latency need addressing, but investment in quantum research and LLM efficiency can overcome these. Limitations include hardware constraints, data requirements, and the need for diverse clinical validation. Future work will refine the system.

## 6 CONCLUSIONS

The prototype presented in this study offers a tangible advancement in drug repurposing by combining quantum-inspired feature extraction with LLM–based analytics. By orchestrating high-dimensional omics datasets—such as RNA-seq and microarray gene expression profiles—with detailed clinical information, the system demonstrates a clear capability to prioritize potential therapies for diverse patient populations. Unlike conventional machine learning methods that struggle to handle complex and expansive data, the quantum-enhanced approach excels at discerning subtle patterns in gene expression, ultimately improving classification metrics such as accuracy, F1-score, and area under the ROC curve. The pipeline's ability to integrate QPCA with LLM-driven interpretation highlights the potential for scalable, explainable, and timely solutions in modern healthcare. Though current hardware limitations and computational demands pose practical challenges, ongoing innovations in quantum simulators and AI architectures will likely reduce operating costs and further streamline this approach.

## ACKNOWLEDGEMENTS

We are grateful to Merrimack College for support.